


\input harvmac\skip0=\baselineskip
\input epsf
\noblackbox
\newcount\figno
\figno=0
\def\fig#1#2#3{
\par\begingroup\parindent=0pt\leftskip=1cm\rightskip=1cm\parindent=0pt
\baselineskip=11pt
\global\advance\figno by 1
\midinsert
\epsfxsize=#3
\centerline{\epsfbox{#2}}
\vskip 12pt
{\bf Fig.\ \the\figno: } #1\par
\endinsert\endgroup\par
}
\def\figlabel#1{\xdef#1{\the\figno}}

\lref\fischler{
W. Fischler, unpublished.
}

\lref\GukovYA{
S.~Gukov, C.~Vafa and E.~Witten,
``CFT's from Calabi-Yau four-folds,''
Nucl.\ Phys.\ B {\bf 584}, 69 (2000)
[Erratum-ibid.\ B {\bf 608}, 477 (2001)]
[arXiv:hep-th/9906070].
}

\lref\GiddingsYU{
S.~B.~Giddings, S.~Kachru and J.~Polchinski,
``Hierarchies from fluxes in string compactifications,''
arXiv:hep-th/0105097.
}

\lref\BeckerGJ{
K.~Becker and M.~Becker,
``M-Theory on Eight-Manifolds,''
Nucl.\ Phys.\ B {\bf 477}, 155 (1996)
[arXiv:hep-th/9605053].
}

\lref\PolchinskiSM{
J.~Polchinski and A.~Strominger,
``New Vacua for Type II String Theory,''
Phys.\ Lett.\ B {\bf 388}, 736 (1996)
[arXiv:hep-th/9510227].
}

\lref\SilversteinXN{E.~Silverstein,``(A)dS backgrounds from
asymmetric orientifolds,''arXiv:hep-th/0106209. }
\lref\BrownDD{J.~D.~Brown and C.~Teitelboim,``Dynamical
Neutralization Of The Cosmological Constant,''Phys.\ Lett.\ B {\bf
195}, 177 (1987). } \lref\AbbottQF{L.~F.~Abbott,``A Mechanism For
Reducing The Value Of The Cosmological Constant,''Phys.\ Lett.\ B
{\bf 150}, 427 (1985).} \lref\BrownKG{J.~D.~Brown and
C.~Teitelboim,``Neutralization Of The Cosmological Constant By
Membrane Creation,''Nucl.\ Phys.\ B {\bf 297}, 787 (1988). }
\lref\HawkingMY{S.~W.~Hawking and I.~G.~Moss,``Fluctuations In The
Inflationary Universe,''Nucl.\ Phys.\ B {\bf 224}, 180 (1983).}
\lref\ColemanAW{S.~R.~Coleman and F.~De Luccia,``Gravitational
Effects On And Of Vacuum Decay,''Phys.\ Rev.\ D {\bf 21}, 3305
(1980).} \lref\BoussoXA{R.~Bousso and J.~Polchinski,``Quantization
of four-form fluxes and dynamical neutralization of the
cosmological constant,''JHEP {\bf 0006}, 006
(2000)[arXiv:hep-th/0004134].}
\lref\deAlwisPR{
S.~P.~de Alwis, J.~Polchinski and R.~Schimmrigk,
``Heterotic Strings With Tree Level Cosmological Constant,''
Phys.\ Lett.\ B {\bf 218}, 449 (1989).
}
\lref\psstw{
J.~Preskill, P.~Schwarz, A.~D.~Shapere, S.~Trivedi and F.~Wilczek,
``Limitations on the statistical description of black holes,''
Mod.\ Phys.\ Lett.\ A {\bf 6}, 2353 (1991).}
\lref\jmls{J.~M.~Maldacena and L.~Susskind,
``D-branes and Fat Black Holes,''
Nucl.\ Phys.\ B {\bf 475}, 679 (1996)
[arXiv:hep-th/9604042].}
\lref\PolyakovJU{
A.~M.~Polyakov,
``The wall of the cave,''
Int.\ J.\ Mod.\ Phys.\ A {\bf 14}, 645 (1999)
[arXiv:hep-th/9809057].
}
\lref\FengIF{
J.~L.~Feng, J.~March-Russell, S.~Sethi and F.~Wilczek,
``Saltatory relaxation of the cosmological constant,''
Nucl.\ Phys.\ B {\bf 602}, 307 (2001)
[arXiv:hep-th/0005276].
}
\lref\ChamseddineQU{
A.~H.~Chamseddine,
``A Study of noncritical strings in arbitrary dimensions,''
Nucl.\ Phys.\ B {\bf 368}, 98 (1992).
}

\lref\KachruHE{ S.~Kachru, M.~Schulz and S.~Trivedi, ``Moduli
stabilization from fluxes in a simple IIB orientifold,''
arXiv:hep-th/0201028.
}
\lref\FreyHF{ A.~R.~Frey and J.~Polchinski, ``N = 3 warped
compactifications,'' arXiv:hep-th/0201029.
}
\lref\DysonNT{L.~Dyson, J.~Lindesay and L.~Susskind,
``Is there really a de Sitter/CFT duality,''arXiv:hep-th/0202163.
}
\lref\CrapsII{ B.~Craps, D.~Kutasov and G.~Rajesh,
``String Propagation in the Presence of Cosmological
Singularities,'' arXiv:hep-th/0205101.
}
\lref\gp{
P.~Ginsparg and M.~J.~Perry,
``Semiclassical Perdurance Of De Sitter Space,''
Nucl.\ Phys.\ B {\bf 222}, 245 (1983).}

\lref\gh{
G.~W.~Gibbons and C.~M.~Hull,
``de Sitter space from warped supergravity solutions,''
arXiv:hep-th/0111072.}

\lref\chu{
C.~M.~Hull,
``de Sitter space in supergravity and M theory,''
JHEP {\bf 0111}, 012 (2001)
[arXiv:hep-th/0109213].}
\lref\fone{T.~Banks,
``Cosmological Breaking Of Supersymmetry?,''
Int.\ J.\ Mod.\ Phys.\ A {\bf 16}, 910 (2001).}
      \lref\ftwo{
R.~Bousso,
``Positive vacuum energy and the N-bound,''
JHEP {\bf 0011}, 038 (2000)
[arXiv:hep-th/0010252].}
\lref\per{
P.~Berglund, T.~Hubsch and D.~Minic,
``de Sitter spacetimes from warped compactifications of IIB string
theory,''
arXiv:hep-th/0112079.}
\lref\vn{
K.~Pilch, P.~van Nieuwenhuizen and M.~F.~Sohnius,
``De Sitter Superalgebras And Supergravity,''
Commun.\ Math.\ Phys.\  {\bf 98}, 105 (1985).}
\lref\fre{
P.~Fre, M.~Trigiante and A.~Van Proeyen,
``Stable de Sitter Vacua from N=2 Supergravity,''
arXiv:hep-th/0205119.}
\lref\kallosh{
R.~Kallosh,
``N = 2 supersymmetry and de Sitter space,''
arXiv:hep-th/0109168.}
\lref\chamblin{
A.~Chamblin and N.~D.~Lambert,
``de Sitter space from M-theory,''
Phys.\ Lett.\ B {\bf 508}, 369 (2001)
[arXiv:hep-th/0102159].}
\lref\susskind{
L.~Susskind,
``Twenty years of debate with Stephen,''
arXiv:hep-th/0204027.}
     \lref\jmas{
J.~M.~Maldacena and A.~Strominger,
``Statistical entropy of de Sitter space,''
JHEP {\bf 9802}, 014 (1998)
[arXiv:gr-qc/9801096].}

\lref\StromingerJG{
A.~Strominger,
``The Inverse Dimensional Expansion In Quantum Gravity,''
Phys.\ Rev.\ D {\bf 24}, 3082 (1981).
}

\lref\GatesCT{
S.~J.~Gates and B.~Zwiebach,
``Gauged N=4 Supergravity Theory With A New Scalar Potential,''
Phys.\ Lett.\ B {\bf 123}, 200 (1983).
}
\lref\LindeSK{
A.~D.~Linde,
``Hard art of the universe creation
(stochastic approach to tunneling and baby universe formation),''
Nucl.\ Phys.\ B {\bf 372}, 421 (1992)
[arXiv:hep-th/9110037].
}
\lref\RohmAQ{
R.~Rohm,
``Spontaneous Supersymmetry Breaking In Supersymmetric String Theories,''
Nucl.\ Phys.\ B {\bf 237}, 553 (1984).
}

\lref\DasguptaSS{
K.~Dasgupta, G.~Rajesh and S.~Sethi,
``M theory, orientifolds and G-flux,''
JHEP {\bf 9908}, 023 (1999)
[arXiv:hep-th/9908088].
}

\lref\KachruGS{ S.~Kachru, J.~Pearson and H.~Verlinde,
``Brane/flux annihilation and the string dual of a
non-supersymmetric  field theory,'' arXiv:hep-th/0112197.
}

\lref\raph{R. Bousso, ''Adventures in de Sitter space'',
hep-th/0205177.}

\lref\raphb{
R.~Bousso, O.~DeWolfe and R.~C.~Myers,
``Unbounded entropy in spacetimes with positive cosmological constant,''
arXiv:hep-th/0205080.
.}

\lref\GarrigaEF{ J.~Garriga and A.~Vilenkin, ``Recycling
universe,'' Phys.\ Rev.\ D {\bf 57}, 2230 (1998)
[arXiv:astro-ph/9707292].
}

\def\a{\alpha}
\def\p{\partial}
\def\half{{1\over2}}
\def\apm{{\a^\prime}}

\Title{\vbox{\baselineskip12pt\hbox{hep-th/0205316}\hbox{SLAC-PUB-9228}
\hbox{HUTP-02/A019}
}}{de Sitter Space in Non-Critical String Theory}

\centerline{Alexander Maloney\footnote{$^\dagger$}{Department of
Physics, Harvard University, Cambridge, MA 02138}, Eva
Silverstein\footnote{$^*$} {SLAC and Department of Physics,
Stanford University, Stanford, CA 94309} and Andrew
Strominger$^\dagger$\footnote{}{\it To appear in Proceedings of
Stephen Hawking's 60th Birthday Symposium} }

\vskip .3in \centerline{\bf Abstract} {Supercritical string theories in
$D>10$
dimensions with no moduli are described, generalizing the
asymmetric orientifold construction of one of the authors
\SilversteinXN. By taking the number of dimensions to be large and
turning on fluxes, dilaton potentials are generated with
nontrivial minima at arbitrarily small cosmological constant
and D-dimensional string coupling,
separated by a barrier from a flat-space linear dilaton
region, but
possibly suffering from strong coupling problems.
The general issue of the decay
of a de Sitter vacuum to flat space is discussed. For relatively
small barriers, such decays are described by gravitational
instantons. It is shown that for a sufficiently large potential
barrier, the bubble wall crosses the horizon.
At the same time the instanton decay time exceeds the Poincare
recurrence time. It is argued that the inclusion of such instantons is
neither physically meaningful nor consistent with basic principles such as
causality. This raises the possibility that such de Sitter vacua are
effectively  stable.
In the case of the supercritical flux models, decays to
the linear dilaton region can be forbidden by such large barriers,
but decays to lower flux vacua including AdS minima nevertheless
proceed consistently with this criterion. These models provide
concrete examples in which cosmological constant reduction by flux
relaxation can be explored. } \vskip .3in

\smallskip
\Date{}
\listtoc
\writetoc


\newsec{Introduction}

Recent progress in string theory has led to deep conceptual
insights into the quantum nature of a number of spacetime
geometries, including black holes and AdS. dS (de Sitter) has so
far been largely left out of the fun.  A key reason for this is
that so far no fully satisfactory  dS solution of string theory
has been found.\foot{ However there are a number of interesting
constructions which may not have been fully exploited
\refs{\GatesCT\chamblin \kallosh  \gh \chu \per -\fre }.} The
problem is intrinsically difficult because there can be no
unbroken supersymmetry in dS \vn . Hence the solutions are likely
to be isolated with no massless scalars or moduli.

A recent approach \SilversteinXN\ employs supercritical
superstring theory. Although they do not have flat space as a
solution, noncritical string theories are of intrinsic interest
for a wide variety of reasons. They are implicated in tachyon
decay processes in compact closed string backgrounds \deAlwisPR,
and in attempts to obtain the QCD string \PolyakovJU. Their
precise place in the M-theory duality web remains an outstanding
question.  New cosmological solutions (with a strongly coupled
singularity) of supercritical string theory were discussed in
\CrapsII.  The recent application to de Sitter space
\SilversteinXN\ utilizes an asymmetric orientifold construction in
non-critical 12-dimensional string theory which has no moduli. The
supercriticality introduces a leading-order cosmological term
(dilaton potential) which aids in fixing the dilaton.  By turning
on RR fluxes it is possible to arrange for the dilaton to have a
nontrivial minimum with a positive cosmological constant. The
string coupling at the minimum is numerically, but not
parametrically, small.
However, as stressed
in \SilversteinXN, the true expansion parameter about the minimum
-- and the nature of string perturbation theory about a minimum
which balances dilaton tadpoles from noncriticality against RR
fluxes -- are not understood.  For both of these reasons the
existence of a string perturbation expansion about the minimum is
in question, and strong coupling effects could in principle
eliminate the dS solution. A second issue in this model is that the dS minimum
is unstable to decay to flat space. This implies that not
every point on the asymptotic boundary of the space is dS. One of the
recent lessons of string theory is that the nature of the boundary can be
quite important, so a theory which asymptotically decays to flat space may
be very different from a ``stable'' dS.

In this paper we report on work in progress which
improves on this construction. A generalized
asymmetric orientifold construction is introduced with a new
parameter: the number of dimensions $D$. By making the number of
dimensions large and employing the Bousso-Polchinski mechanism \BoussoXA\
with the RR fluxes we are able to make the cosmological constant
at the minimum parametrically small, the higher-dimensional string
coupling parametrically  weak, and the effective barrier to the linear
dilaton regime parametrically large.
Despite this improvement we have not
understood the true expansion parameter about the minimum, which
could therefore in principle be eliminated by strong coupling
effects.

In particular, as a function of the dimensionality $D$, the number
of RR fields is $n_{RR}=2^D$, which dominates the spectrum at
large $D$.  This is potentially both a liability and an asset: on
the one hand, the $2^D$ RR species threaten to render the
effective coupling uncontrollably large; on the other hand, the
large number of RR fluxes facilitate the construction of vacua
with small cosmological constant and weak $D$-dimensional string
coupling.  As one increases $D$, the naive number of degrees of
freedom increases, and as we will see one can obtain a larger and
larger de Sitter space. It is tempting to speculate that the $2^D$
RR degrees of freedom pertain to the entropy; this will be
interesting to explore in the future.
In particular, since a large de Sitter space
requires a large number of states (to account for the large
entropy), the large number of degrees of freedom intrinsic to
supercritical string theory may play a natural role.

We also consider, in a more general setting, the issue of the decay
of dS space to flat space. When the barrier is small such decays
clearly occur via flat space bubble nucleation and are described
by gravitational instantons. However, the required bubble size
grows with the barrier height, and eventually the bubble wall
crosses the horizon. We will argue that the inclusion of such
superhorizon processes has bizarre consequences. Causality and
unitarity appear to be  violated, and for very large height the
process describes the tunneling of the entire universe to a
planckian region! The proper rules for dS quantum gravity are not
well understood, and this casts doubt on the assertion that such
instantons should be included in the first place. We further
note that the tunneling time exceeds the
Poincare recurrence time for dS \DysonNT\ for exactly the same parameter
range that the instanton becomes superhorizon sized. (It
also exceeds the (shorter) time for all of de Sitter space to
tunnel into a maximal black hole \gp.) Hence both the observable
significance and the validity of the semiclassical approximation
are in question for the superhorizon decay processes.\foot{As discussed in
section 3.4 and alluded to in \FengIF,
this is a de Sitter analog of the breakdown of the semiclassical
approximation for black holes discussed in \psstw. Related discussions can
be found in \refs{\raph,\raphb}.}   If the
superhorizon instantons are excluded, a ``false" dS vacuum may be
stable against decay to flat space (or to the linear dilaton
regime in the case of the supercritical models), or equivalently
the decay time may
become so long as to be meaningless.

In the supercritical models, one can in this way potentially
forbid decays from a large range of dS minima to the linear
dilaton regime, since as we will see the domain wall tension is
too large for a sub-horizon size bubble. However, we also find
decays between different flux vacua proceeding via nucleation of
D-branes (as in \refs{\BrownKG,\BoussoXA,\FengIF}), including
transitions from dS to AdS. The model thus is a stringy
construction sharing features with those studied in
\refs{\AbbottQF,\BrownKG,\BoussoXA,\FengIF} exhibiting a dynamical
relaxation of the cosmological constant. Among the different flux
vacua, there are many more choices of flux configuration yielding
larger values of the cosmological constant than smaller values,
and in our system there are large degeneracies among different
flux vacua due to the highly symmetric structure of the internal
dimensions.

This paper is organized as follows. Section 2.1 presents the
asymmetric orientifold construction. 2.2 describes the de Sitter
minima, and 2.3 discusses the lower limit on the cosmological
constant implied by flux quantization.  3.1 reviews the instantons
which describe the tunneling from de Sitter to flat space.  3.2
questions the conventional wisdom that this tunneling occurs (or
is even well-defined) for arbitrarily high barriers. 3.3 relates this
to Poincare recurrence and the breakdown of the semiclassical
approximation. Finally in
section 3.4 we address the stability of the asymmetric orientifold
models.

\newsec{de Sitter Compactifications of Super-Critical String Theory}

In this section we generalize the construction of \SilversteinXN\
to large numbers $D$ of dimensions and describe de Sitter
solutions of the low energy action. We compute the contributions
to the dilaton potential from noncriticality, orientifold planes
and RR fluxes. We demonstrate that by taking the number of
dimensions to be large, one can find potentials having minima at a
parametrically small value of the $D$-dimensional string coupling.
Finally, we consider flux quantization and show that at large $D$
the cosmological constant can be made parametrically small.


\subsec{Asymmetric Orientifolds in Non-Critical String Theory}

In D (more than 10) dimensions, we start with the string frame low
energy effective theory for the graviton, dilaton and Ramond-Ramond fields
\eqn\act{ S_D = {1\over2\kappa_D^2}\int d^Dx
\sqrt{-G}\left(e^{-2\phi}\left(R-{2(D-10)\over3\a'} + 4 \nabla_\mu\phi
\nabla^\mu \phi\right)-\half \sum_p(F_p)^2\right) } where the sum runs
over the various RR fields $F_p$ in the theory.

We will be interested in asymmetric orientifold models obtained
from this $D$-dimensional theory in which the dilaton is fixed.
Let us begin by noting a few salient points regarding the spectrum
in these relatively unfamiliar theories. Note from the action
\act\ (and as discussed in \ChamseddineQU\ and reviewed in
\SilversteinXN), the graviton, dilaton, and RR fields in $D$
dimensions are massless.  However, if one calculates using free
field theory the putative zero-point energy of these fields in
flat (string-frame) space, i.e. in the linear dilaton background,
one finds in the NS sector a vacuum energy of $-(D-2)/16$.  As
explained in \ChamseddineQU, this reflects the effective tachyonic
behavior of the fields in the linear dilaton background (obtained
from \act\ by expanding in small fluctuations about the linear
dilaton solution).  In order to obtain the effective mass squared
of the fields in the Lagrangian expanded around a putative
extremum with constant dilaton (such as those we are studying in
this paper) one must therefore cancel the contribution from the
linear dilaton from the zero point energy.  This amounts to the
statement that in the NS sector, the effective vacuum energy $E$
is off from the free field result $E_0$ by \eqn\enshift{
E=E_0+{{D-10}\over{16}} . }

Let us now proceed to the models of interest here, which are
compactifications from $D$ down to $d=D-r$ dimensions. We will
eventually be interested in the case of large $D$ with $d$ held
fixed, and in particular how various quantities depend on $D$.
Because as we will see the quantities relevant to our conclusions
scale exponentially with $D$, some numerical factors which are
order one will not be explicitly computed.

We begin with a self-dual torus $T^r$. The zero modes on the torus
are given by \eqn\pconventions{\eqalign{ &
p^i_L={1\over\sqrt{\apm}}(m^i+n^i)\cr &
p^i_R={1\over\sqrt{\apm}}(m^i-n^i)\cr }} and the dimensions of
the corresponding worldsheet operators are $({\apm \over 4} p_L^2,
{\apm \over 4} p_R^2)$. Mod out by the orientifold group generated
by \eqn\genI{ g_1\equiv (0,s^2)_{d+1}\dots (0,s^2)_{d+r} }
\eqn\genII{g_2\equiv (-1,1)_{d+1}\dots (-1,1)_{d+r}} \eqn\genIII{
g_3\equiv \Omega I_r} \eqn\genIV{g_4\equiv (-1)^F (s,s)_{d+1}\dots
(s,s)_{d+r}}
As in \SilversteinXN, we adopt the following notation.
$(0,s^2)_i$ is an asymmetric shift on the $i^{\rm th}$ coordinate, and
acts as $(-1)^{n^i+m^i}$. $(s,s)_i$ is a geometric shift on the $i^{\rm
th}$ coordinate by
half the circle radius, and acts as $(-1)^{m^i}$.
$\Omega$ is an
orientation reversal, $I_r$ a reflection on all $r$ coordinates of
the $T^r$. $(-1,1)_i$ is a reflection on the $i^{\rm th}$ left-moving
coordinate only, and is
at the heart of the moduli-fixing effect of this model, since it
projects out all the untwisted NS NS moduli.

In order to check level-matching (for modular invariance) and to
check for twisted moduli, we must compute the vacuum energy in all
inequivalent sectors, taking into account \enshift.
Let us start with the shifts.  In the $(0,s^2)^r$ twisted sector,
the momentum and winding lattice \pconventions\ is shifted so that
$(m,n)\to (m+1/2,n+1/2)$, while in the $(s,s)$ sector it
is shifted by $(m,n)\to (m, n+1/2)$.  Each $(0,s^2)$ shift (per direction)
has a right-moving energy of $1/4$, while each  $(s,s)$ shift (per
direction)
gives left and right moving energies of $1/16$.
For the element $g_2=(-1,1)^r$, we have ground state energies
\eqn\enI{
(E_L={r\over 8}-\half, E_R=-\half)
}
This level-matches if $r=4k$ for integer $k$.  In order to avoid any
massless modes (potential moduli) in this sector, we must take $k>1$.
For the element $g_2g_4$
we have
\eqn\enIIIV{
(E_L={r\over 8}-\half={{k-1}\over 2}, E_R={r\over 16}-\half={{k-2}\over
4}),
}
requiring that $k\equiv 2N$ be even for level-matching.
As discussed in \ChamseddineQU, in order to have a standard GSO
projection,
one requires $D=d+r\equiv 8j+2$ for integer $j$.
Altogether,
in order to have a consistent orientifold group we need
\eqn\consI{
r=8N
}
for integer $N\ge 1$, and to have an ordinary GSO projection we need
\eqn\consII{
d=D-8N=8j+2-8N
.}

This model has two sets of orientifold planes -- O-$(d-1)$-planes
generated by the element $g_3$ and spacefilling O-$(D-1)$-planes
generated by the T-dual element $g_2 g_3 g_2=\Omega$. We also have
anti-orientifold planes, which are necessary to cancel the RR
tadpoles -- these are generated by the elements $g_3 g_4$ and $g_2
g_3 g_4 g_2$.
The total contribution to the action due to these orientifold planes is
\eqn\sor{
S_{\rm Orientifold} = \sum_{i} T_{O_i} \int d^{p_i+1}x \sqrt{-G} e^{-\phi}
}
where $i$ runs over the orientifolds -- here
the orientifold
group acting on the $r$ dimensions of our torus introduces $2^{r-1}$
$O$-$(d-1)$ planes, $2^{r-1}$ $\bar O$-$(d-1)$ planes,
as well as the T-dual objects, an
$O$-$(D-1)$ plane, and an $\bar O$-$(D-1)$ plane.
These T-dual pairs are identified under the action of $g_2$,
 so \sor\ is just $2^r$ times the action for a single $O_{d-1}$
plane: \eqn\asdII{\eqalign{ S_{\rm Orientifold} &=  2^r T_{O_{d-1}}
\int d^dx \sqrt{-G} e^{-\phi} \cr
    &= -{2^{7/2+D/4}\pi^{1/2} \over \kappa_D
\ell_s^{1-D/2+d}} \int d^dx \sqrt{-G_d} e^{-\phi}. }} Here we have
defined the string length \eqn\dls{\ell_s=2\pi\sqrt{\apm}} and are
using the generalized formula for the tension of an orientifold
$p$-plane in $D$ dimensions derived in \SilversteinXN\ with the
assumptions listed there (which consist essentially of the
procedure \enshift\ for the closed-string channel modes applied to
the annulus diagram),
\eqn\top{ 2^{D-(p+1)}T_{O_p} = -{2^{7/2+D/4}\pi^{1/2}\over
\kappa_D\ell_s^{p+2-D/2}}. }

The action of the orientifold group projects out the NS-NS moduli
of the $T^{r}$, so the $d$-dimensional action for the untwisted
NS-NS sector reduces to \eqn\asd{\eqalign{ S_{\rm NS}= {1\over 2
\kappa_d^2 } \int d^dx\sqrt{-G_d}e^{-2\phi} \left(R_d - {2(D-10)
\over 3 \apm}+4\nabla_\mu\phi \nabla^\mu\phi\right) }} where the
$d$-dimensional gravitational coupling is \eqn\asd{ \kappa^2_d=
{\kappa^2_D\over v\ell_s^{r}} = v^{-1}\ell_s^{d-2} .} $v$ here is
the dimensionless effective volume of the compactification space
given by \eqn\asd{\biggl(\int_{T^{r}}
d^{4m}x\sqrt{-G_{r}}\biggr)_{eff}= v \ell_s^{r},} and is of order
one.  We have taken the $D$-dimensional coupling to be
$\kappa^2_D=\ell_s^{D-2}$ . \foot{In making this choice, we are
tacitly assuming that high order terms in the perturbation series
will be $\le$ order one with respect to this choice of coupling.
See the discussion in
\SilversteinXN\ for more details.}

Note that one could also consider multiple copies of this
orientifold group acting on subtori of $T^r$. Each $\Omega I_p$
action reduces the RR spectrum by half, so this has the virtue of
reducing the number of species which contribute to the effective
coupling. However, there is a danger of also reducing the effective
volume and thus $v$ in \asd, thereby increasing the effective
coupling.  It would be interesting to determine the winner of the
competition between these two effects, but for now we will stick
to a single copy of the orientifold group \genI -\genIV.

We now turn on some RR fluxes along the compact directions
(see, e.g. \refs{\DasguptaSS\KachruHE\FreyHF\GukovYA\GiddingsYU\BeckerGJ
{--}\PolchinskiSM}). 
In $D$ dimensions, a $p$-form field strength wrapped on a cycle of
volume $V_p$ will be quantized as \eqn\asdfl{
{1\over2\kappa_D}\int_{V_p} F_p = {\sqrt{\pi}} \ell_s^{2p-D\over2}
Q } where $Q$ is an integer.  Let us use a basis of cycles given
by the square subtori $\subset T^r$.  We will label these by
$i=1,\dots,2^r$.  Turning on RR fluxes adds a dilaton independent
piece to the $d$ dimensional string frame action. Before
orientifolding, there are $r \choose k$ possible $k$-form fluxes
to choose from, for a total of $2^{r}$. Although some of the
internal fluxes will be projected out by the orientifold action,
certain flux configurations will be left invariant. These
invariant combinations of fluxes from the untwisted sector of the
orbifold, which involve fluxes of different rank related to each
other by T-duality, will also be subject to the quantization
condition inherited from the parent theory.
Because our orbifold is of finite order independent of $r$, the number
of invariant fluxes still scales like $2^r$ for large $r$ after
taking into account the reduction in the RR spectrum effected
by the orientifold action.  Chern-Simons couplings
among the many RR fields at large $D$ may also affect the
spectrum in a given flux background, and the set of
consistent choices of flux configuration; this would be interesting
to work out in detail.

Going to $d$-dimensional Einstein frame \eqn\eins{ G_d{}_{\mu\nu}\to
\tilde
G_{\mu\nu} =G_d{}_{\mu\nu}e^{4\phi\over 2-d} }
the low energy action becomes \eqn\asd{ S= {1\over2\kappa_d^2}\int
d^dx \sqrt{-\tilde G} \left(\tilde R - \left({4\over
(d-2)}\right)\p_\mu\phi\p^\mu\phi- {1\over
v\ell_s^{2}}U(\phi)\right).
} The Einstein frame dilaton potential is \eqn\pot{ U(\phi) =
e^{{4\over d-2} \phi}(a - b e^{\phi} + c e^{2\phi}) } where
\eqn\abc{\eqalign{ a&=v 4\pi^2 \left({2(D-10)\over3}\right) \cr
b&= 2 \ {2^{7/2+D/4}\pi^{1/2}}v_O \cr c&= \sum_{i=1}{\pi \over
v_{p_i}} Q_i^2 \ell_s^{2p_i-D}+\Lambda_1 \equiv \pi
\sum_{i=1}{\tilde Q_i^2}+\Lambda_1 .}} Here in the expression for
$b$, $v_O$ is a dimensionless volume associated with the
orientifold planes on our orbifold similar to $v$; again this is
of order 1 in our model and we will not keep track of such factors
in our analysis. In the expression for $c$, $i$ labels the fluxes
in the square basis discussed above, and we consider only
invariant combinations of these basic fluxes. $p_i$ is the degree
of the field strengh and $v_{p_i}$ is an order one dimensionless
volume associated to the $i^{\rm th}$ flux. (Before the
orientifolding, these volumes are self-dual, but as in
\asd
the effective volumes may
be reduced by the action of the orientifold group.)

$\Lambda_1$ is the one-loop dilaton potential.  It will be
proportional to $n_{RR}\sim 2^D$ times $\xi(D)$ where $\xi(D)$ is
an unknown $D$-dependent constant, which is related to the
effective loop-counting parameter in our theory.  (For some
insight into the scaling of loop effects in gravitational field
theory as a function of dimension $D$, see \StromingerJG, where
factors of $1/D!$ appear with additional loops, providing enhanced
control at large $D$.) Because the $2^D$ RR bosons dominate the
spectrum, $\Lambda_1$ is likely to be negative in the string
theoretically regulated theory, similarly to the situation in for
example Scherk-Schwarz compactifications \RohmAQ\ and many other
non-supersymmetric orbifold examples that have been analyzed in
critical string theory, in which one finds the sign of $\Lambda_1$
to be the same as that of the difference between the number of
massless fermions and bosons in the tree-level spectrum.  Below,
we will analyze the potential assuming conservatively that
$\Lambda_1\sim -2^D$ for definiteness, but as will become clear
the qualitative results apply for a large range of possible values
of $\Lambda_1$ including those with smaller magnitude.

In principle, we should also include a renormalization of Newton's
constant at the same order; this will not affect the
perturbative stabilization in what follows in this section,
but nonperturbatively
may adjust the instanton actions in \S3.

\subsec{de Sitter Solutions}

Let us write the potential as
\eqn\eid{\eqalign{
U(\phi) &= \biggl(a -be^{\phi}+{b^2\over4a}(1+\delta)e^{2\phi}\biggr)
e^{{{4\phi}\over{d-2}}}.
}}
There is a de Sitter solution if $U(\phi)$ has
a stable minimum at positive energy.
This requires that the solutions of $U'(\phi)=0$
\eqn\gst{
e^{\phi_\pm} ={a\over db}
\biggl({{d+2\pm\sqrt{(d-2)^2-8d\delta}}\over 1+\delta}\biggr)
}
are real -- here $\phi_\pm$ is the local minimum (maximum).
In addition the effective cosmological constant
\eqn\poten{\Lambda=U(\phi_+),}
should be greater than zero.  These two conditions require that
\eqn\crit{0<\delta<{{(d-2)^2}\over 8d}.}
As $\delta$ increases from the lower bound to the upper bound, $U(\phi_+)$
increases from $0$ to
${{a^{{{d+2}\over{d-2}}}8^{4\over d-2} (d-2)^2}\over
{b^{4\over d-2}d(d+2)^{{{d+2}\over d-2}}}}$.
and the string coupling decreases from ${2 a\over b}$ to
${{8a}\over{b(d+2)}}$.
Near $\delta=0$, the cosmological constant goes like
\eqn\kss{U(\phi_+)=a({2a\over b})^{4\over d-2}\delta+ {\cal O}(\delta^2).}

If we wish to minimize the string coupling we must take
$\delta\sim {{(d-2)^2}\over 8d}$.  For example,
in the original scenario of \SilversteinXN\ ($D=12$, $d=4$) this gives
\eqn\critu{\eqalign{
a={8\pi^2\over3},~~~~~~b={\pi^{1/2}2^{15/2}},~~~~~~
\Lambda \sim {1\over(2\pi \alpha^{1/2})^4} (0.05),
~~~~~~ e^{\phi_+}\sim 0.11.
}}
This has the disadvantage that $\Lambda$ is only a couple orders of
magnitude above
string scale.  Also, the potential barrier seperating the local
minimum from the
global minimum at $\phi\to-\infty$ is small, so the vacuum is not very
stable against
tunneling effects.
Let us instead try to minimize $\Lambda$ by taking $\delta\to0$.
We find that (modulo issues of flux quantization, which we will consider
in the next section) we can make $\Lambda$ as small as we like, with
\eqn\critl{
\Lambda \sim 0, ~~~~~~ e^{\phi_+}\sim 0.16.
}
We have found that $\Lambda$ can be made arbitrarily small, at the cost of
a small
increase in the string coupling.
In addition, this solution is much more stable, since the potential
barrier is high.

A solution with small D-dimensional string coupling is found by
taking $a/b\to0$.  From the expressions \abc\ it is clear that
this can always be accomplished by taking $D$ large. However, it is
not clear that this implies a small true effective string coupling
after compactification. The latter may for example be enhanced by
the enormous multiplicity ($\sim 2^D$) of RR fields.  (On the other
hand, if things work as in \StromingerJG, there may in fact be
overcompensating loop-suppression factors as a function of $D$
that preserve the smallness of the effective coupling.)

\subsec{Solutions With Small $\Lambda$}

In order to get a small cosmological constant we must take
$\delta\to0$. However, flux quantization constrains how small we
can get $\delta$, and thus how small we can get $\Lambda$. We see
from \S2.2\ that for $\Lambda\sim 0$ and large $D$, $c$ approaches
a large value \eqn\flu{ c = \pi \sum_i\tilde Q^2_i +\Lambda_1 \to
{b^2\over4a}\sim {2^{D/2}\over 4a} } For example, this is
${3072\over\pi}$ in the scenario in \SilversteinXN. Since
$\Lambda_1\sim -2^D$, we have \eqn\radI{ \pi\sum_i\tilde
Q_i^2\sim 2^D. }
By taking linear combinations of many different fluxes we can tune
$c$ quite accurately -- this is similar to the mechanism of Bousso
and Polchinski \BoussoXA, though in our case we have large
degeneracies in the set of flux configurations. The allowed
charges $Q_i$ lie on a $q\sim 2^r\sim 2^D$-dimensional lattice.
Because of the flux quantization condition, the smallest jumps we
can have in $c$ are of order 1. Because of \radI\ and the fact
that we have $2^D$ independent fluxes $\tilde Q_i$ to pick, there
will always be some $\tilde Q_i$ which are of order 1 (or
smaller), so order 1 jumps are indeed possible. Using \kss, this
gives for the scale of the lowest-lying de Sitter minima
\eqn\changec{\eqalign{ &\Delta c\sim {b^2\over 4a}\delta\sim 1\cr
& \Lambda = U(\phi_+) \sim \left( {a\over b}\right)^{2d\over d-2}\sim
2^{-{Dd \over 2(d-2)}} }}
Since $b\sim 2^{D/4}$, this vacuum energy is exponentially small
for large $D$.


\newsec{Metastability of the de Sitter Vacuum}

In addition to the de Sitter minimum, the dilaton potential \pot\
has a global minimum with vanishing cosmological constant at
$\phi\to-\infty$. Our system also has a multitude of different dS
and AdS vacua obtained from different configurations of flux in
the internal space. This raises the issue of whether or not the de
Sitter minimum is only metastable. This question arises
generically in any string construction of a de Sitter solution
involving a potential which vanishes at weak coupling, and/or
containing many flux vacua.

Instantons have been described \refs{\ColemanAW\AbbottQF,\BrownKG}
which might be related to this tunneling. However, as we will see
in this section, when the barrier between the minima is
sufficiently large, the instanton degenerates and no longer
describes tunneling of a de Sitter horizon volume to a comparably
sized-region of flat space. The instanton describes a rather
unphysical process in which the visible universe disappears
altogether. Such ``super-horizon" instantons occur in the
parameter range for which the bubble wall lies behind the horizon.

   Whether or not such processes actually occur, and whether or not
such de Sitter vacua can be stable,  are questions which cannot
be definitively settled with our present understanding of quantum
gravity in de Sitter space. In ordinary field theory, instantons
provide saddle point approximation to a functional integral with
fixed boundary conditions. The instantons which describe the
decay/disappearance of de Sitter space have no boundary at all,
and so it is not clear if they should be included. We will argue
that the super-horizon instantons in a sense violate both
causality and unitarity and should be omitted altogether. We will
also discuss other potential mechanisms for mediating vacuum
decay.

\subsec{The Instantons}
    For simplicity we work in the thin wall approximation, in which
case the relevant instanton solutions are rather simple. They have
been described in detail in \BrownKG\ and will now be reviewed.

The euclidean solutions are characterized by the tension $T$ of
the bubble wall and the dS cosmological constant $\Lambda$. The
solutions are determined by simply matching the extrinsic
curvatures on the two sides of the bubble wall to the tension $T$
in accord with the Israel junction condition. The instanton looks
like a portion of a round sphere glued to a portion of flat space.
The spherical portion is
\eqn\sprp{ds^2=R_{dS}^2\bigl(d\theta^2+sin^2\theta
d\Omega_3^2\bigr), ~~~~~~0\le\theta \le \arcsin {R_B \over
R_{dS}},} where $d\Omega_3^2$ is the metric on the unit three
sphere, $R_{dS}=\sqrt{3 / \Lambda}$ is the dS radius, and
$R_B$ is the radius of the $S^3$ boundary. The flat space portion
is \eqn\fsp{ds^2=dr^2+r^2d\Omega_3^2,~~~~~~ 0 \le r \le R_B.} The
full instanton is then obtained by gluing together \sprp\ and
\fsp\ along the $S^3$ bubble wall at radius $R_B$. This is
depicted in figure 1a-c. The Israel junction condition
\eqn\lblv{{1 \over R_B^2}={1 \over R_{dS}^2}+\bigl({1 \over
TR_{dS}^2 \kappa^2}-{T \kappa^2\over 4}\bigr)^2} where $M_P$ the Planck
mass,
determines $R_B$ in terms of $T$.  Note that $R_B$ increases with
$T$ for small $T$ but then decreases for $T$ greater than the
critical value \eqn\tcrt{T_C={2 \over \kappa^2 R_{dS}} .} $R_B$
approaches zero for very large $T$.

\fig{
The Euclidean instanton solutions matching the sphere (Euclidean de
Sitter)
to flat space.
The cases $T<T_C$, $T=T_C$ and $T>T_C$ 
are shown in figures {\bf a)}, {\bf b)} and {\bf c)} respectively.
}{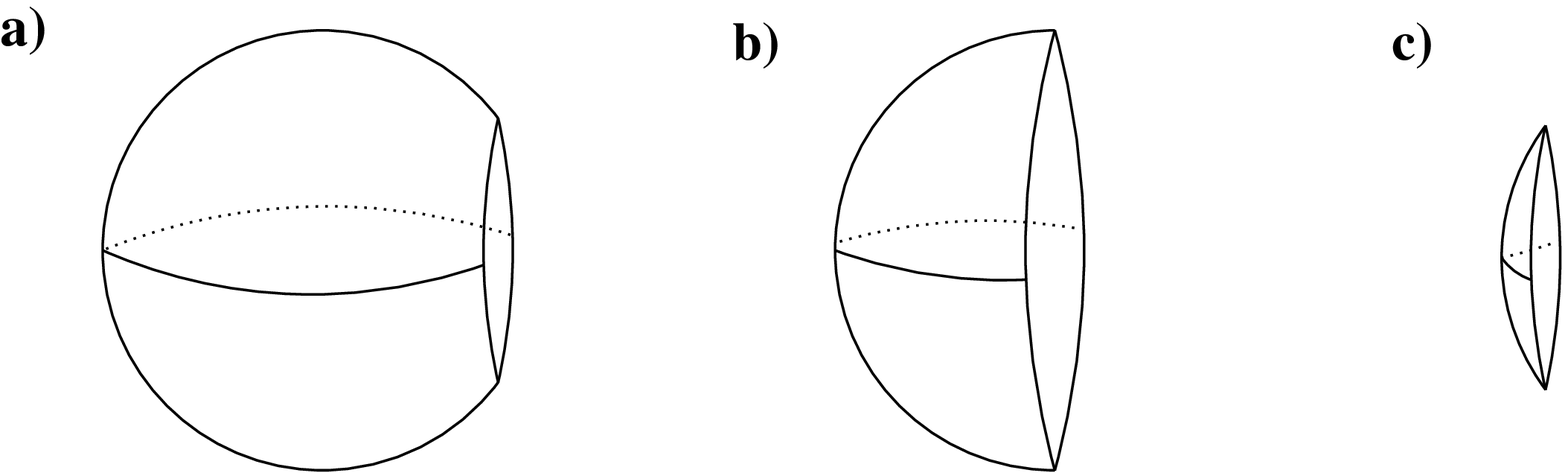}{4in}

It is straightforward to generalize these euclidean solutions to the
dS$\to$dS and dS$\to$AdS cases.
In general dimension $d$, the relation \lblv\ becomes \BrownKG\
\eqn\genR{
{1\over R_B^2}={2\Lambda_o\over (d-2)(d-1)}+
\biggl({\kappa_d^2T\over 2 (d-2)}+{\Lambda_i-\Lambda_o\over (d-1)T
\kappa_d^2}
\biggr)^2.
}
Here, $\Lambda_o$ is the initial dS cosmological constant (outside the
bubble)
and $\Lambda_i$ is the final cosmological constant $\Lambda_i$ (inside the
bubble).
In general dimension $d$, the critical tension is
\eqn\tcrtgen{T_C^2={2(\Lambda_o-\Lambda_i)(d-2)\over (d-1) \kappa_d^4}.}

The instanton purportedly describes tunneling from one classical
geometry to another. We are interested in an initial dS geometry.
The final geometry is then given by the analytic continuation of
the instanton, which describes an expanding bubble of flat space
inside dS. The two geometries are glued together along the moment
of time symmetry. This is depicted in figure 2a-c. The tunneling
rate is purportedly given by the action of the instanton minus the
background action of Euclidean dS without a bubble.
This is
\eqn\iact{
\Delta S = 2\pi^2 R_B^3 T +
{2\pi^2 \over R_{dS}^2\kappa^2}
\left[
2 R_{dS}^4 \mp \left\{3 R^3_{dS} (R_{dS}^2-R_B^2)^{1/2} - R_{dS} (R_{dS}^2
- R_B^2)^{3/2} \right\}
\right]
.}
The upper and lower signs correspond to $T<T_C$ and $T>T_C$, respectively.
Again, the expression for general $d$ was worked out in \BrownKG\
(equations (6.4)-(6.7)).

\fig{
The Lorentzian instanton geometry describing the nucleation of a
bubble of flat space (the shaded region) inside de Sitter space.
The cases $T<T_C$, $T=T_C$ and $T>T_C$ 
are shown in figures {\bf a)}, {\bf b)} and {\bf c)} respectively.
}{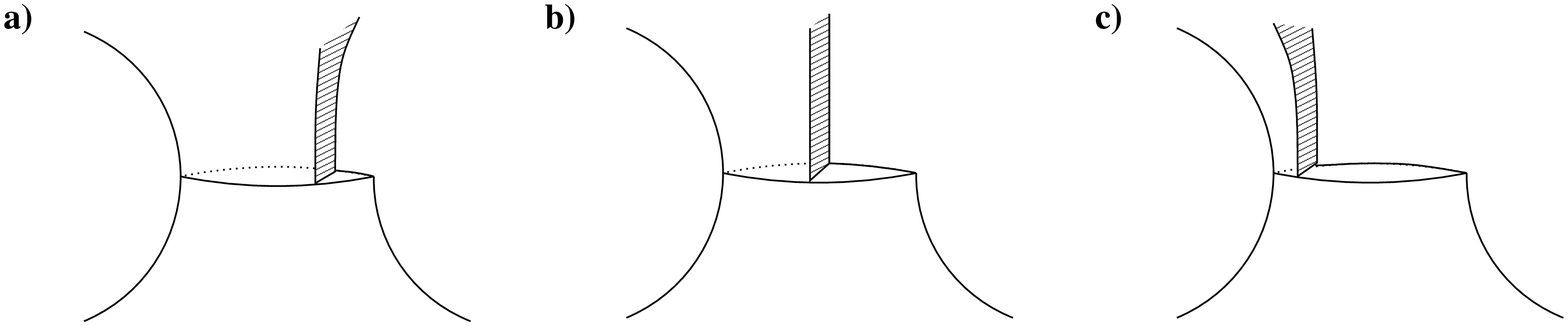}{6in}


\subsec{Causality}

The tunneling process depicted for small tensions in figure 2a
approaches the usual flat space false vacuum decay in the limit
$M_P \to \infty$ with $T$ held fixed. The rate according to \iact\
also approaches the correct flat space value.  The instanton of
figure 1a surely describes this tunneling process for sufficiently
small but finite $T\over M_P^3$.

The process depicted in figure 1c on the other hand has a bizarre
interpretation.  The entire universe tunnels to a small dime, with
one flat and one dS face! Furthermore for $T \to \infty$ the rate
from \iact\ approaches a constant. Hence the tunneling rate can be
enhanced by adding a large number of ultra-planckian domain walls.
In fact, the action \iact\ is not monotonically increasing in the
regime $T>T_C$; for certain ranges of paramaters, the tunneling
rate {\it increases} as the tension increases!  This conflicts with
the notion of decoupling in low-energy field theory, as well as
the general fact that tunneling effects are supressed as the size
of the barrier increases.

This process also appears in conflict with causality. An observer
in dS should be insensitive to any physics behind the horizon. In
particular there should be no consequences of placing boundary
conditions on the fields along a  timelike surface behind the
horizon. It is easy to find boundary conditions that forbid the
super-horizon instanton. Therefore the observer can learn about
physics behind the horizon by waiting to see whether or not the
tunneling occurs.

There is also an issue with unitarity. In the benign process of
figure 2a, an observer at the south pole finds him or herself,
after the tunneling,  in the middle of a bubble of flat space.
However for the superhorizon case of figure 2c, his or her entire
southern causal diamond - the entire observable universe -
disappears.  It has been advocated by many (see for example \jmas\
and the contribution \susskind\ to these proceedings) that the
causal diamond should be viewed as a closed unitary system (whose
microstates compute the entropy). Surely this process - in which
the diamond disappears altogether- violates unitarity in the worst
possible manner!

Based on these observations, our conclusion is that when the
tension $T$ exceeds $T_C$,  the superhorizon instantons simply
should not be included in the semiclassical description of dS. At
the same time we wish to stress that, with our current level of
understanding of dS quantum gravity, no such conclusions can be
drawn with certainty.

The above arguments apply equally well to tunneling from de Sitter
to de Sitter or Anti-de Sitter, with the critical tension given by
\tcrtgen.  We should note that the criterion \eqn\tlt{T>T_C} in
the case $\Lambda_o=0$, $\Lambda_i<0$ reproduces the well known
Coleman and DeLuccia condition for the stability of flat space
against tunneling to Anti de Sitter \ColemanAW. We may thus regard
the stability criterion \tlt\ as a generalization of the Coleman -
DeLuccia mechanism.

Even if such instantons are not to be included, there may be other
processes which mediate the decay of the dS to flat space when the
barrier is very high. For example if de Sitter space is viewed as
a thermal ensemble\foot{with temperature conjugate to the energy
defined by the timelike Killing vector which preserves the causal
diamond}, thermal fluctuations could eventually push the value of
$\phi$ over the top (see e.g. \LindeSK\GarrigaEF\ for a discussion
of this mechanism). This however is also not obviously possible.
There appears to be a maximum energy allowed in dS given by the
largest black hole which can fit inside the observer horizon. If
the energy required to cross the barrier to flat space  exceeds
this value, it may be suppressed. Furthermore if the appealing
notion \refs{\fone,\ftwo,\fischler} that dS has a finite number of states
given by the area law is accepted, there must be a highest energy
state. Again if this is less than the barrier height decay to flat
space is suppressed.

\subsec{Breakdown of the Semiclassical Approximation}
There is yet another way to interpret the condition
$T^2>T_C^2 \sim \Lambda$,
which involves a further assumption about de Sitter quantum gravity.
Following \fone, we assume that de
Sitter gravity has a finite number of degrees of freedom which determine
the de Sitter quantum entropy.
Imagine in this context a  detector sitting on a timelike geodesic for a
very long time.    The  detector must be built out of a subset of
the finite number
of degrees of freedom, all of which will eventually be thermalized by
de Sitter radiation from the horizon.  This thermalization process sets a
maximum timescale in de Sitter space, intervals longer than which can
never be measured by a geodesic observer. (See also \refs{\raph, \raphb}.)
The precise value of the thermalization time depends on the structure of
the detector, but it is certainly
less than the Poincare recurrence time,
which is a timescale on which {\it all} degrees of freedom have been
thermalized.
This recurrence time is related to the de Sitter entropy by \DysonNT\
\foot{The authors of \DysonNT\ considered several different types of
recurrence phenomena.
Here we quote the timescale for two point fluctuations proportional the
thermal background value of the Green function -- so called ``relative"
fluctuations --
as opposed to fluctuations of some fixed size independent of $S$.}
\eqn\prt{\eqalign{
t_{\rm recurrence} \sim \exp\{S\} =
\exp\left\{ {8\pi^2 R_{dS}^2 \over \kappa^2}
\right\}
.}}

Another time scale in de Sitter is the typical time for the entire space
to
tunnel to a maximal sized black hole. This has been estimated using
instantons in \gp\ as
\eqn\gpr{t_{\rm black hole}\sim \ (t_{\rm recurrence})^{1/3}.}
Hence the entire space tunnels into a maximal black hole exponentially
many times before the Poincare recurrence time.

We wish to compare these times
to the expected lifetime of de Sitter space due to
vacuum decay.
When the tension equals the critical value $T_C$, the lifetime
for the putative instanton decay is
(omitting a prefactor which is polynomial in $R_{dS}$)
\eqn\lifet{\eqalign{
t_{\rm decay}\sim &\exp\{\Delta S\}=\exp\left\{ {8\pi^2 R_{dS}^2 \over
\kappa^2}
\right\}
.}}
This is precisely the Poincare recurrence time!
Thus as $T$ approaches the critical value $T_C$
the lifetime becomes comparable to the recurrence time, and no observer
will ever live long enough to see the vacuum decay.
\foot{Given that the action \iact\ {\it decreases} at $T>>T_C$ one might
worry that naively
applying the instanton methods for very large tensions
would lead to decay timescales shorter than the recurrence time.
This turns out not to be the case: as $T\to\infty$ the decay time
precisely approaches \lifet.
}
Moreover, at
$T=T_C$ the lifetime is much longer than the time \gpr\
\eqn\asd{\eqalign{
t_{\rm decay} &\sim t_{\rm recurrence} \cr
            &\sim t^3_{\rm black \ hole}.
}}
Hence in order to observe the decay of de Sitter space when $T=T_C$
one needs a detector capable of passing through a black hole
exponentially many times. We regard the existence of such detectors
doubtful!

Let us state this in yet another way. The semiclassical approximation
describes the de Sitter horizon as a hot wall in contact with a heat
reservoir with $infinite$ heat capacity.  In this approximation
no correlations ever appear in
the radiation emitted from the horizon. In the exact theory, it is
plausible that the horizon has a finite heat capacity as determined
from the finite de Sitter entropy. This means that if we watch long enough
correlations will be seen in the radiation.\foot{Of course, as mentioned
above, no one can live that long. However this only underscores the
unphysical nature of a tunneling process which takes such a long time.}
A typical  time required to see those correlations is the Poincare recurrence time.
Hence  this time scale signals the breakdown of the semiclassical
approximation. A semiclassical instanton which involves a longer time
scale therefore cannot be trusted.

Phrased in this way, our argument parallels a similar one give for
black holes in \psstw, and alluded to in the de Sitter context in \FengIF.
In \psstw, it was argued that the semiclassical approximation
for near-extremal black holes breaks down as the temperature goes
to zero very near extremality.  The breakdown occurs when the
energy of a typical thermal Hawking quantum exceeds the
excitation energy of the black hole above extremality. Clearly the
Hawking emission cannot proceed under these circumstances
 because it would leave a
subextremal black hole with a naked singularity.

This is a close analogy to
the situation we have described in the de Sitter context.
The hot horizon emits a thermal spectrum of  bubbles of flat space.
When the energy of these bubbles (as determined in part by the tension of
the bubble walls) exceeds the energy of de Sitter space above flat space,
the semiclassical approximation breaks down.

In the black hole case, it was eventually quantitatively understood \jmls\
in the context of string theory that this breakdown of the semiclassical
approximation signals the appearance of a gap. Presumably similarly
interesting and yet-to-be understood phenomena appear in the
de Sitter context.

In conclusion, superhorizon tunneling processes from dS to flat space
do not appear to be  meaningful or consistent.
The stability and correct quantum description
of a dS vacuum separated by a very
high barrier from flat space is an open question.

\subsec{Instantons in the Orientifold Model}


In the asymmetric orientifold model the tension of the domain wall
separating the de Sitter from the flat vacuum at $\phi\to-\infty$
is determined by the shape of $U(\phi)$; for example in $d=4$ it
is roughly \eqn\tens{T\sim {a^{3/2}\over b}.}
Using the criterion of the previous subsection, we conclude that
many of the de Sitter minima discussed in section 2 are stable
against decay to the linear dilaton regime. \foot{ We should note
that when $D$ is large, the thin wall approximation breaks down
for the potentials \pot; in this limit the width of the domain
wall interpolating between the de Sitter and flat vacua scales as
$T^{-1}$. However, this subtlety does not affect the causality
considerations of Section 3.2. } The maximum-energy de Sitter
minimum stable under this
decay 
is at $c\sim {b^2\over 4a}+{\cal O}(b^2)$, i.e. at $\delta\sim 1$.  (Here
we are only keeping track of exponential dependence on $D$, i.e. factors
of $b$ but not $a$.)
This corresponds to an energy of the order
\eqn\enmax{
U_+^{max}\sim {1\over b^2}\sim 2^{-D/2}
}
The minimum-energy de Sitter minima possible with our quantization
condition on the charges and thus on $c$ (which are of course
also stable under this decay)
have $c\sim {b^2\over 4a} + {\cal O}(1)$, i.e. at $\delta\sim {a\over
b^2}$.
This corresponds to an energy of the order
\eqn\enmin{
U_+^{min}\sim {1\over b^4}\sim 2^{-D}.
}



In addition to the instanton decays to the linear dilaton regime
discussed above, there is also the possibility of transitions
among the different flux vacua, as in
\BrownKG\BoussoXA\FengIF\KachruGS.  D-branes extended along $d-1$
of the $d$ de Sitter dimensions constitute domain walls separating
vacua with different flux configurations. More specifically,
D-branes of charge $Q$ connect vacua of flux $Q_1$ and $Q_1-Q$ on
the dual cycle to the D-brane on the compactification.  In order
to determine the (in)stability of our solutions, we must apply the
results reviewed in \S3.1\ to such D-brane induced decays in
addition to the dilatonic domain wall we considered above.


At our de Sitter minima for $d=4$, the string coupling is
\eqn\mincoup{g_s \sim 1/b} and the energy is \eqn\endS{U_+\sim
(a^4/b^4)(c-(b^2/4a))} As we just discussed, the lowest-lying dS
vacua have $c$ tuned to cancel $b^2/(4a)$ to within order 1, so
that \eqn\enminII{U_+^{min} \sim 1/b^4} The highest-lying dS
minima that are stable against decay to the linear dilaton
background have, from our earlier calculation, $c$ tuned such that
$c-(b^2/4a)\sim b^2$, i.e. \eqn\enmaxII{U_+^{max} \sim a^2/b^2}
Recall from \tcrtgen\ that \eqn\critagain{T_C^2\sim
\Lambda_o-\Lambda_i.} The D-brane tension is, in Einstein frame,
from \SilversteinXN\ and the above scaling of $g_s$ at the
minimum, \eqn\Dten{T \sim (1/b^2) 2^{-D/4}.} This formula will
apply for a transition in which the bubble wall is a single
D-brane; the tension of multiple D-branes will be subject to
appropriate binding energy contributions.

If we allow the instanton, i.e. if $T<T_C$, then its action $B$ is
given by equation (6.4) in \BrownKG.  One should keep in mind that
the renormalization of Newton's constant may affect the overall
scaling of the action.
In addition to the contribution of exp(-action) to the probability
for decay, there will also be significant degeneracy factors from
the large multiplicity of vacua in our large-$D$ system.
Here we will confine ourselves to checking whether the transitions
occur at all according to the criterion we have developed in this
paper, assuming that the semiclassical instanton analysis applies
(i.e. that the action is large enough in renormalized Planck
units).

For example, consider decays from $U_+^{max}\to
U_+^{max}-a^4/b^4$. This occurs if $c\sim\sum (\tilde Q_i)^2$
changes by order 1, and in particular can proceed via a bubble
consisting of a single D-brane . In this case, the D-brane tension
is \eqn\TenI{T_{(i)}\sim 2^{-D/4} (1/b^2),} while the critical
tension in this case is \eqn\crittenI{T_{C(i)}\sim 1/b^2 .} So
$T_{(i)}<<T_{C(i)}$, and the decay proceeds according to our
criterion developed above.


Similarly, there are decays from dS to AdS. Consider for example a
transition $U_+^{min}\to -U_+^{min}$. Here again
\eqn\tenIII{T_{(iii)}\sim 2^{-D/4} 1/b^2} and
\eqn\crittenIII{T_{C(iii)}\sim 1/b^2} so the decay is again
allowed.


As we mentioned above, there will be large factors in the
transition rates associated with the relative multiplicity of
different decay endpoints. In particular, the smaller the value of
$\sum \tilde Q_i^2\equiv R^2$ coming into the coefficient $c$, the fewer
choices of flux configuration there are in the window
between $R$ and $R+\Delta R$ for a fixed $\Delta R$.  So
although decays to AdS are possible, it is reassuring that this
degeneracy factor prefers the less negative $\Lambda_i$ values.
(In fact these factors also prefer higher dS vacua to lower ones, which
may act to suppress the decays depending on the scaling of the
renormalized instanton action.)


 \centerline{\bf Acknowledgements} We are grateful to
M. Aganagic, T. Banks, R. Bousso, S. Kachru, A. Karch, A. Linde,
S. Minwalla, L. Motl, L. Susskind, N. Toumbas and A. Vilenkin for 
useful conversations.
This work was supported in part by DOE grant DE-FG02-91ER40654 and
under contract DE-AC03-76SF00515 and by the A.P. Sloan Foundation.

\listrefs
\end